\documentclass[sigconf]{acmart}

\usepackage{amsmath}
\usepackage{amsfonts}
\usepackage{bm}
\usepackage{booktabs}
\usepackage[inline]{enumitem}
\usepackage{subcaption}
\usepackage{color}
\usepackage{xcolor}

\AtBeginDocument{%
  }

\begin{document}

\title{How do Large Language Models Understand Relevance? \\ A Mechanistic Interpretability Perspective}

\author{Qi Liu}
\email{qiliu6777@gmail.com}
\affiliation{%
  \department{Gaoling School of Artificial Intelligence}
  \institution{Renmin University of China}
  \city{Beijing}
  \country{China}
}

\author{Jiaxin Mao}
\email{maojiaxin@gmail.com}
\authornote{Corresponding author.}
\affiliation{%
  \department{Gaoling School of Artificial Intelligence}
  \institution{Renmin University of China}
  \city{Beijing}
  \country{China}
}

\author{Ji-Rong Wen}
\email{jrwen@ruc.edu.cn}
\affiliation{%
  \department{Gaoling School of Artificial Intelligence}
  \institution{Renmin University of China}
  \city{Beijing}
  \country{China}
}

\begin{abstract}

Recent studies have shown that large language models (LLMs) can assess relevance and support information retrieval (IR) tasks such as document ranking and relevance judgment generation. However, the internal mechanisms by which off-the-shelf LLMs understand and operationalize relevance remain largely unexplored. In this paper, we systematically investigate how different LLM modules contribute to relevance judgment through the lens of mechanistic interpretability. Using activation patching techniques, we analyze the roles of various model components and identify a multi-stage, progressive process in generating either pointwise or pairwise relevance judgment. Specifically, LLMs first extract query and document information in the early layers, then process relevance information according to instructions in the middle layers, and finally utilize specific attention heads in the later layers to generate relevance judgments in the required format. Our findings provide insights into the mechanisms underlying relevance assessment in LLMs, offering valuable implications for future research on leveraging LLMs for IR tasks.\footnote{Our code is public at \url{https://github.com/liuqi6777/llm-relevance}.}

\end{abstract}



\maketitle

\section{Introduction}


\emph{Relevance} is one of the most important concepts in information retrieval~\cite{saracevic1975relevance}. It serves as the cornerstone for determining the usefulness of retrieved information in response to a query, bridging the gap between user intent and document representation. Traditionally, relevance has been understood through explicit feedback mechanisms, such as user judgments, and operationalized in ranking algorithms to optimize retrieval effectiveness.


With the advent of large language models (LLMs), such as GPT-4~\cite{openai2023GPT4}, the notion of relevance has undergone a paradigm shift. These models, trained on vast corpora and equipped with sophisticated contextual understanding, bring new dimensions to relevance assessment in information retrieval. Specifically, LLMs do not rely solely on traditional relevance signals, such as term frequency, but instead leverage deep semantic embeddings and latent representations to assess relevance. In relation to specific tasks, relevance judgment and document ranking represent two of the most prevalent examples of utilizing LLMs for assessing relevance. The former necessitates that LLMs assign a relevance score to a document in response to a query and specific annotation criteria~\cite{faggioli2023Perspectives, abbasiantaeb2024Can}, whereas the latter mandates that LLMs rank several documents according to their relevance to the given query~\cite{sun2023ChatGPT, qin2023Large}.

While LLMs demonstrate promising performance in these tasks, the underlying mechanisms by which they understand and operationalize the concept of relevance remain opaque. Unlike traditional systems such as BM25~\cite{robertson2009Probabilistic}, where features and weights are explicitly defined, LLMs derive relevance judgment from complex interactions within layers of neural architecture. These interactions are shaped by pretraining and fine-tuning processes, making it challenging to pinpoint how relevance signals are internally encoded and utilized in downstream tasks, including document ranking and relevance annotation. Since most existing LLMs are built on the transformer architecture, it is crucial to investigate the roles and contributions of token embeddings, the multi-head attention mechanism, multilayer perceptrons, and intermediate activations in relevance assessment.

In this paper, we aim to address a critical question: \emph{how do large language models understand and operationalize the concept of relevance?} 
We want to know: 
(i) whether there exist some common components in LLMs that can encode relevance signals that are independent of specific tasks (i.e., effective across various tasks such as relevance judgment or ranking, as well as diverse prompt formats);
and
(ii) how these relevance signals are conveyed within the LLMs during the forward pass.

\begin{figure*}[t!]
    \centering
    \includegraphics[width=0.99\textwidth]{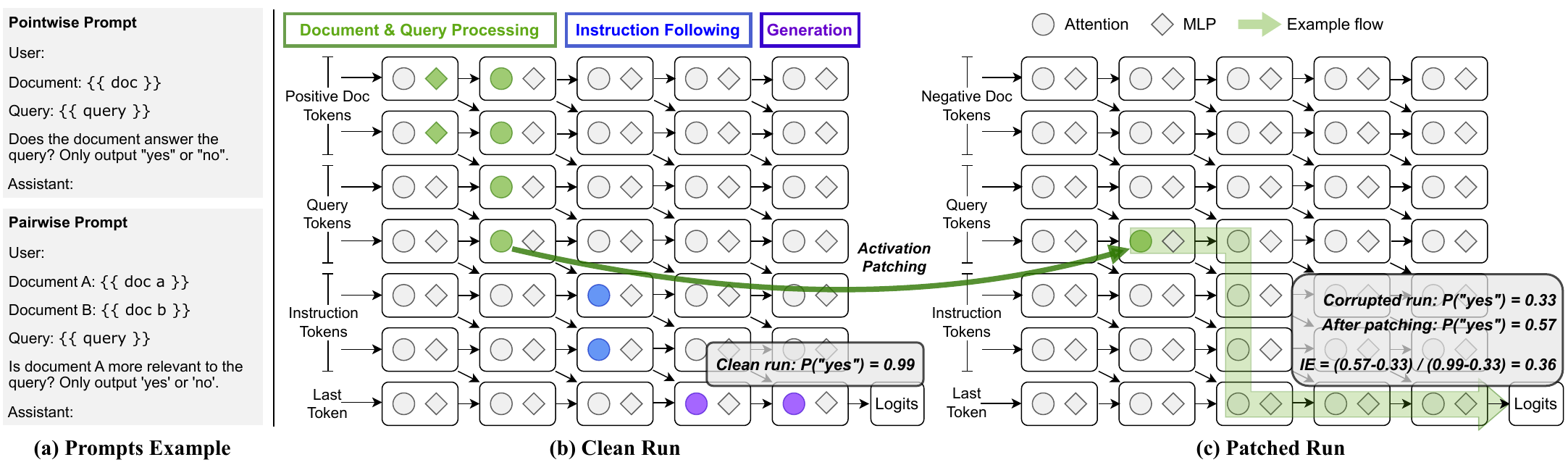}
    \caption{Visualization of an activation patching example using pointwise style prompt (shown in (a)). Activation patching computes the effect of a specific module by running the LLM three times: a clean run (b) with the positive document, a corrupted run with the negative document, and a patched run with corrupted input but the activation of the selected module is replaced with the value in the clean run. Then the effect is computed based on the patched output. Potential information flow within LLMs is shown in (b): LLM first capture information in document and query in early layers (\textcolor[HTML]{60A917}{green} modules), then receive task information in middle layer (\textcolor[HTML]{0000FF}{blue} modules), finally control the result generation in deeper layer (\textcolor[HTML]{6600CC}{purple} modules).}
    \label{fig:overview}
\end{figure*}

To investigate these questions, we adopt methods from \emph{Mechanistic Interpretability}, which seeks to understand the internal mechanisms of language models, with a particular focus on \emph{activation patching}~\cite{vig2020Investigating, meng2022Locating}. Activation patching, also known as causal mediation analysis~\cite{vig2020Investigating}, causal tracing~\cite{meng2022Locating}, or interchange intervention~\cite{geiger2021Causal}, allows us to isolate and manipulate specific model components to trace how useful signals flow through the network. 
Specifically in this paper, we aim at interpreting the behavior and mechanisms of LLMs on some simple relevance judgment tasks using two forms of prompts (e.g., pointwise prompt ``\textit{Does the document answer the query?}'' or pairwise prompt ``\textit{Is the first document more relevant than the second to the query?}''). 
We hypothesize that the relevance signals are carried out by a specific subset of the network. Then, we test this hypothesis by adopting activation patching. Figure 1 illustrates how we use activation patching techniques to analyze relevance judgment by an LLM. By selectively replacing activations in certain layers with those from other contexts, we can measure the causal impact of individual components on the output of these two relevance-related tasks. This granular analysis provides insights into the hierarchical and distributed nature of relevance processing in LLMs.

We conduct a set of analysis experiments to track the information flow of relevance signals within the model and identify the components that play a role in relevance judgment. 
Our findings suggest that LLMs encode relevance as a multi-faceted construct, distributed across different layers and modulated by the interaction between attention heads and feed-forward networks. 
Preliminary results indicate that LLM processes and transmits information progressively, specifically,  
(i) A group of MLPs and attention heads in early layers captures information from documents and queries, offering potential relevance signals;
(ii) Middle layers receive task information from instructions and interact between the instruction and the previous document and query;
(iii) A few attention heads in the latter half layers are used to control the output format.
We also find that LLMs have the same mechanism and share the same functional circuits under pointwise and pairwise prompts, which supports the hypothesis that LLMs use a specific subset of the network to operationalize relevance. 

We further evaluate the mentioned components on downstream tasks using the knockout technique~\cite{wang2022Interpretability}. When knocking out these components, LLMs lose their effectiveness in relevance judgment and document ranking, which demonstrate the necessity. We verify the above findings across different LLMs and datasets.

In summary, the main contributions of this paper are as follows:
\begin{itemize}
    \item We first use the activation patching techniques to explain how LLMs understand relevance internally.
    \item We discover the mechanism of LLMs processing and transmitting information in relevance judgment.
    \item We verify through experiments that our findings could be applied to different models and datasets.
\end{itemize}

\section{Related Work}

\subsection{Assessing Relevance Using LLMs}

Relevance assessment has long been a central task in information retrieval, manifested in typical downstream tasks such as relevance judgment and document ranking. 
Recent studies have explored using LLMs for relevance judgment, where LLMs are prompted with labeling guidelines and expected to generate a relevance label for each query-document pair that aligns with human assessors~\cite{faggioli2023Perspectives, thomas2023Large, abbasiantaeb2024Can}. Some studies endeavor to generate automatic labels for the entire dataset utilizing LLMs~\cite{abbasiantaeb2024Can}, while others concentrate on effectively addressing gaps within the evaluation pool~\cite{abbasiantaeb2024Can, macavaney2023OneShot}.

Different from relevance judgment, document ranking does not require LLMs to generate explicit labels but to rank several documents based on their relevance to a query and return an ordered list~\cite{sun2023ChatGPT, chen2024TourRank}. \emph{Pointwise} approaches are very similar to relevance judgment, however, they usually only use the likelihood score as relevance score to rank document~\cite{liang2022Holistic, zhuang2023Yes}. \emph{Pairwise} approaches receive two documents and judge which is more relevant~\cite{qin2023Large}, which is also closed to relevance judgment. There are also listwise methods~\cite{sun2023ChatGPT} and selection-based methods~\cite{chen2024TourRank} that take multiple documents as inputs. 

Although approaches based on LLMs exhibit remarkable performance, the mechanisms through which LLMs encode and operationalize relevance remain opaque, necessitating further investigation into their inner workings. In this paper, we will focus on interpreting relevance judgment with simple pointwise and pairwise style prompts.

\subsection{Interpretability in Information Retrieval}

Interpretability has become a focal point in IR research, as understanding how models make decisions is critical for improving trustworthiness~\cite{anand2022Explainable}. 
With pre-trained language models like BERT~\cite{devlin2019BERT}, however, the interpretability challenge becomes more complex due to their high dimensionality and non-linearity. Existing work includes analyzing the internal features of the model, for example, usually using probing methods to analyze the internal representation of the BERT-based model (e.g., embeddings or attention patterns) to reveal whether they have learned certain features or concepts~\cite{zhan2020Analysis, wallat2023Probing, liu2023Understanding, macavaney2022ABNIRML, formal2021Match}.
More recently, ~\citet{chowdhury2024Probing} employed probing to analyze the features within the LLM-based rankers.
These approaches provided insights into how models allocate importance to input features but often fall short of explaining the deeper causality within the model.

\subsection{Mechanistic Interpretability}

Mechanistic interpretability seeks to clarify the behaviors of machine learning models, usually neural networks, by comprehending the underlying algorithms employed by these models~\cite{elhage2021mathematical, olah2017feature}. Typical research directions in mechanistic interpretability include Sparse Auto-encoders~\cite{cunningham2023Sparse, elhage2022Toy}, LogitLens~\cite{nostalgebraist2020interpreting}, and circuit discovery~\cite{wang2022Interpretability}.
Recent advances in mechanistic interpretability have emphasized dissecting the network into interpretable components or circuits. Activation patching, also know as causal tracing~\cite{vig2020Investigating} or causal mediation analysis~\cite{meng2022Locating, geiger2024Causal}, in particular, is a standard tool to isolate and manipulate intermediate activations to trace the flow of information within a model~\cite{zhang2024Besta}. This paradigm has been employed by studies to decode the precise pathways of information flow within models when handling different tasks~\cite{wang2022Interpretability, stolfo2023Mechanistic, olah2020Zoom, meng2022Locating, nanda2023Progress}. In the information retrieval field, \citet{chen2024Axiomatic} employed activation patching to interpreting dense retrieval models. Compared to previous work, we are focusing relevance assessment with LLMs.

\section{Preliminaries}

In this section, we will introduce some background knowledge, including the definition and notation of the Transformer model, the tasks used in this paper, and the specific process of activation patching in this task.

\subsection{Transformer Models}

We restrict our scope to large language models that are based on auto-regressive Transformer architecture~\cite{vaswani2017Attention}, and we describe the basic architecture below using notation similar to~\citet{elhage2021mathematical}.

Generally, assume an input sequence $X = t_1, ..., t_N$ of $N$ tokens. Each token $t_i$ is then embedded as a vector $\bm{x}^{(0)}_i \in \mathbb{R}^d$ using an embedding matrix $\bm{W}_E \in \mathbb{R}^{|\mathcal{V}| \times d}$, over a vocabulary $\mathcal{V}$. Then the input embeddings are processed through a sequence of $L$ transformer layers, each consisting of multi-head attention (MHA) and feed-forward network (FFN) that is usually a multilayer perception (MLP), and a residual stream is applied between each module. Formally, for each layer $l \leq L$, the representation $\bm{h}^{(l)}$ of one token $t$ (we will ignore the foot script $i$ below for convenience) is computed from the previous layer:\footnote{For brevity, layer normalization is omitted as it is not essential for our analysis.}
\begin{equation}
\label{eq:layer}
    \bm{h}^{(l)} = \bm{h}^{(l-1)} + \bm{a}^{(l)} + \bm{m}^{(l)},
\end{equation}
where $\bm{a}^{(l)}$ and $\bm{m}^{l}$ are the output of the MHA module and the FFN module at layer $l$, respectively. More specifically, the output of the MHA module inside the $l$-th layer can be decomposed into the sum of the outputs of each independent attention head: $\bm{a}^{(l)} = \sum_{j=1}^H \bm{a}^{(l, j)}$, $\bm{a}^{(l, j)}$ is the projection of the output of the $j$-th attention head (out of $H$ heads) and is obtained by follows:
\begin{equation}
\label{eq:mha}
    \bm{a}^{(l, h)} = \bm{A}^{(l, h)}\left( \bm{X}^{(l-1)} \bm{W}^{(l, h)}_V \right) \bm{W}^{(l, h)}_O,
\end{equation}
\begin{equation}
\label{eq:attn_pattern}
    \bm{A}^{(l, h)} = \mathrm{Softmax}\left( \frac{\left( \bm{X}^{(l-1)} \bm{W}^{(l, h)}_Q \right) \left( \bm{X}^{(l-1)} \bm{W}^{(l, h)}_K \right)^T}{\sqrt{d/H}} \right),
\end{equation}
where $\bm{X}^{(l)}$ is a matrix with all token representations at layer $l$, $\bm{W}^{(l, h)}_Q, \bm{W}^{(l, h)}_K, \bm{W}^{(l, h)}_V \in \mathbb{R}^{d \times \frac{d}{H}}$ are projection matrices and $\bm{W}^{(l, h)}_O \in \mathbb{R}^{\frac{d}{H} \times d}$ are the output matrix for the $h$-th attention head at layer $l$, and $\bm{A}^{(l, h)}$ is the attention patterns.
In the end, output logits are obtained from the final layer representations via a prediction head:
\begin{equation}
\label{eq:logit}
    \mathrm{Logits} = \bm{W}_U \bm{x}^L + \bm{b} \in \mathbb{R}^{|\mathcal{V}|}.
\end{equation}

In this paper, we study which layers and modules of the LLM, i.e., $\bm{a}^{(l)}$ and $\bm{m}^{(l)}$, contribute to the relevance judgment and how they work.

\subsection{Prompting LLMs to Relevance Judgment}

We consider the task of relevance judgment in this paper. Specifically, we adopt two different prompting strategies in the experiments, namely \emph{pointwise} and \emph{pairwise}. 

For the pointwise strategy, the prompt evaluates the relevance of a single document to a given query. This method directly assesses whether the content of a document is relevant to the query, as illustrated in the upper part in Figure~\ref{fig:overview}(a). By instructing the LLM to provide binary outputs (``yes'' or ``no''), we ensure simplicity and clarity in relevance judgment.
In contrast, the \emph{pairwise} strategy focuses on comparing two documents in terms of their relevance to the same query. This approach, illustrated in the lower part in Figure~\ref{fig:overview}(a), asks the LLM to determine which document is more relevant. The pairwise prompt enables a relative comparison between two documents, which is particularly useful in ranking tasks~\cite{qin2023Large}.
To align with the pointwise prompt and simplify the analysis procedure, we also limiting the response of LLMs to binary labels, instead of instructing them to directly output the identifier of the more relevant document.

Both strategies leverage the natural language understanding capabilities of LLMs, but they differ in their applicability. The pointwise prompt is suitable for binary relevance classification, while the pairwise prompt is more appropriate for evaluating ranked relevance. By investigating the similarities and differences in the internal mechanisms of large language models when using these two different prompts, we can obtain a comprehensive understanding of how LLMs assess relevance in both the relevance judgment task and document ranking task.


\subsection{Activation Patching}

\begin{figure}[t]
    \centering
    \includegraphics[width=0.9\linewidth]{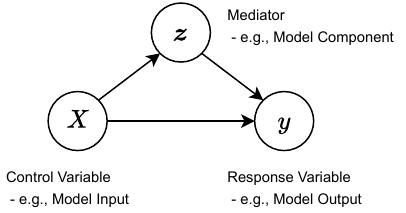}
    \caption{Illustration of Causal Mediation Analysis.}
    \label{fig:cma}
\end{figure}

Activation patching identifies critical components of a model by intervening in their latent activations. It's also known as causal mediation analysis~\cite{vig2020Investigating}, which aims to measure how a treatment effect is mediated by intermediate variables~\cite{pearl2001direct}. As illustrated in Figure~\ref{fig:cma}, activation patching consider specific component $z$ within the model as intermediaries in the causal path from inputs $X$ to outputs $y$, and then the causal effect of $z$ on $Y$ can be measured.

Specifically, this methodology involves a clean prompt (denoted as $X_{\text{clean}}$) along with a corrupted prompt (denoted as $X_{\text{Corrupted}}$, and encompasses three distinct model runs: a \emph{clean run} on $X_{\text{clean}}$, a \emph{corrupted run} on $X_{\text{Corrupted}}$, and a \emph{patched run} on $X_{\text{Corrupted}}$ with a specific model component's activation restored from the cached value of the clean run.

A significant challenge in activation patching lies in the construction of prompt pairs. In prior studies, corrupted prompts were typically generated by altering specific keywords, such as substituting nouns in clean prompts~\cite{wang2022Interpretability}. This approach is more applicable to straightforward fact-based judgments or knowledge tasks; however, it is inadequate in the context of relevance assessment, where queries and documents can vary considerably. In such cases, merely modifying a few words is insufficient to fundamentally alter the relevance of a document. 

To address this issue, we adopt a strategy that for a given query, we entirely replace the positive document with a negative one to construct a data pair. Formally, for each query $q$, we construct one positive passage $d_{pos}$ and one negative passage $d_{neg}$, and denote each data point as a triplet $(q, d_{pos}, d_{neg})$. In a pointwise prompt, we use positive document in clean run and use negative document in the corrupted run, while in a pairwise prompt, in the clean run we put the positive document before the negative document and vice versa in the corrupted run. The full prompts are shown in Figure~\ref{fig:overview}(a). 

In this case, in both pointwise and pairwise prompts, the expected output of the clean run is ``yes'', while that of the corrupted run is ``no''. When performing activation patching, we can observe the output after patching to find out which modules in the model can restore the model behavior in the clean run, that is, make the model output ``yes'' again.

We can patch activations into the LLM in different components and different token positions, which can provide us with a more fine-grained analysis of model behavior. For model components, we can patch the activation of the attention outputs (i.e., $\bm{a}^{(l)}$), individual attention heads (i.e., $\bm{a}^{(l, j)}$), or MLP outputs (i.e., $\bm{m}^{(l)}$). 
As for token positions, in the relevance judgment task of this paper, we select the following positions in the complete prompt for patching: 
(1) all positions of the documents (one document for pointwise while two for pairwise); 
(2) all positions of the query; 
(3) all positions of the instruction; 
and (4) the position of the last token.

To summarize, the full activation patching procedure is shown as follows:
\begin{enumerate}[label=\arabic*.]
    \item Construct a prompt pair, including a clean prompt $X_\text{clean}$ and a corrupted prompt $X_\text{corrupted}$ that cause the difference in model behavior.
    \item Choose the model component (e.g., MLP output) and token position (e.g., the last token) to patch.
    \item \emph{Clean run}: run the model on $X_\text{clean}$ and cache activations of the selected component.
    \item \emph{Corrupted run}: run the model on $X_\text{corrupted}$ and record the model outputs.
    \item \emph{Patched run}: run the model on $X_\text{corrupted}$ with a specific activation restored from the cached value of the clean run, and see how the model output has changed compared with the corrupted run.
\end{enumerate}

\paragraph{Metrics}

To measure the specific patch’s causal effect on model output, we first compute the logit difference as a metric:
\begin{equation}
    \mathrm{LD} = \mathrm{Logits}(\text{``yes''}) - \mathrm{Logits}(\text{``no''}),
\end{equation}
and the indirect effect (IE) is defined as
\begin{equation}
    \mathrm{IE} = \frac{\text{LD}_{\text{patched}} - \text{LD}_{\text{corrupted}}}{\text{LD}_{\text{clean}} - \text{LD}_{\text{corrupted}}}, 
\end{equation}
where the subscripts are different runs. Following~\citet{wang2022Interpretability}, we normalize the IE so that it generally falls within the range of $[0, 1]$, where a value of 1 signifies fully restored performance and a value of 0 indicates the performance of the corrupted run. We use the IE to determine which component contributes more to the model behavior.

\section{Experiments}
\label{sec:exp}

In this section, we first describe the experimental setup, and then elaborate on the analysis results.

\subsection{Experimental Setup}
\label{sec:setup}

\paragraph{Models}

We study instruction-tuned LLMs, as they are more commonly employed in document ranking and relevance judgment tasks than foundation models. Specifically, we choose LLama-3.1-8B-Instruct~\cite{grattafiori2024llama3} for most experiments, since it's widely applied and performs well in ranking and relevance judgment. Notably, we didn't fine-tune it using any supervised data, instead, we directly leveraged the off-the-shelf LLM for relevance assessment in a zero-shot manner. We acknowledge that it may see related tasks during instruction tuning, however, our focus here is to investigate how LLMs perform relevance judgment but not how they obtain the ability of relevance judgment during pre-training or post-training. 

Additionally, we also conduct supplementary experiments on other LLMs to explore whether our findings on the Llama can be generalized to different LLMs in Section~\ref{sec:generalize}.

\paragraph{Data}

\begin{figure*}[t]
    \centering
    \includegraphics[width=0.9\textwidth]{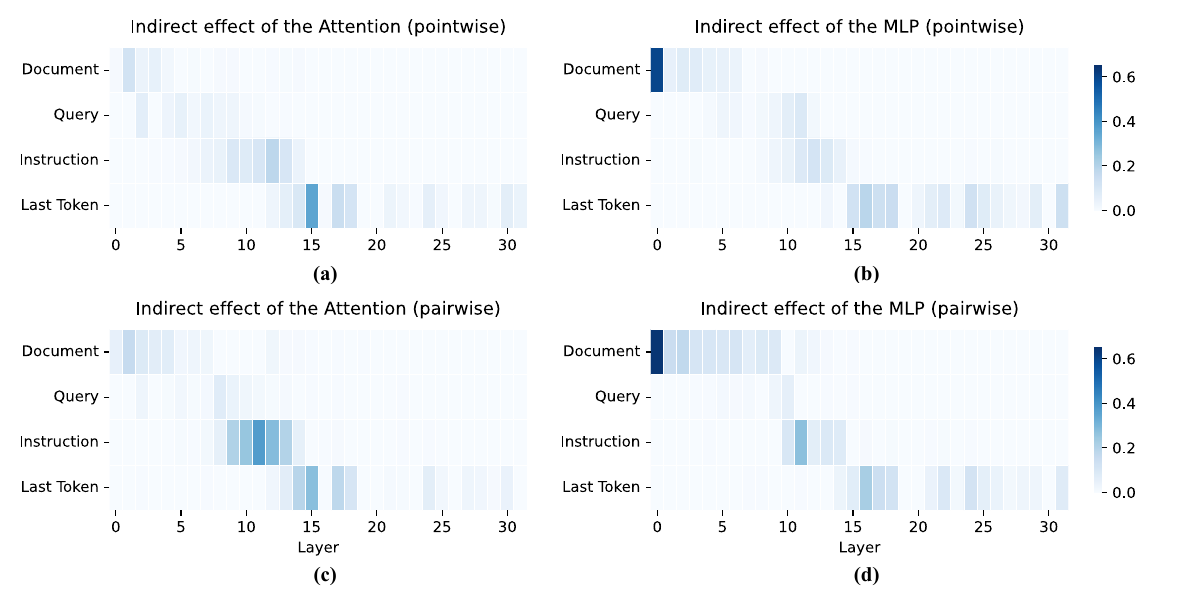}
    \vspace{-3mm}
    \caption{Indirect Effect of different model components at different token positions within LLama-3.1-8B-Instruct on 100 samples from MS MARCO.}
    \label{fig:layerwise-patching}
\end{figure*}

The experiment datasets are conducted from MS MARCO passage ranking dataset~\cite{bajaj2018MS} and Natural Questions (NQ)~\cite{kwiatkowski2019Natural}. For each, we randomly sample 100 queries from the training set and construct one positive document and one negative document for each query. The positives are from the human-labeled training set and we directly use them. The negatives are sampled from the top 100 documents retrieved by BM25~\cite{robertson2009Probabilistic}. To facilitate activation patching, we truncate the data by reducing the length of both documents to that of the shorter document. For each experiment, we compute the average indirect effect over all prompts. We use MS MARCO for most experiments and use NQ to validate that the findings can be generalized across datasets. 

The selection of sample size is based on computational efficiency. For each sample, we need to carry out a complete activation patching process for each component of LLM, including three complete forward passes. This is a relatively large computational cost under the parameter size of the LLM, especially when considering each attention head (for example, there are 1024 attention heads for Llama-3.1-8B). Therefore,  we constructed a dataset of 100 sample levels.

\subsection{Tracing Information Flow}
\label{sec:layer}

We first use activation patching to measure the average indirect effect of the attention block and MLP at different token positions in each layer within LLama-3.1-8B-Instruct. Figure~\ref{fig:layerwise-patching} shows the results of both pointwise and pairwise prompts.

In general, we can find that \textbf{the model processes and transmits information progressively}. As shown in Figure 3, the visual patterns for the indirect effect of both attention layers and MLP layers are similar as they both resemble a diagonal pattern. Notably, there are three main findings: (i) for the initial half of the input corresponding to the document and query, the earlier layers of the model demonstrate a higher indirect effect; (ii) the latter half of the input corresponding to the instruction reveals a significant indirect effect in the middle layers; (iii) in deeper layers (from the layer 15 onwards), the indirect effects in these positions approach zero, while the indirect effect at the last token position begins to increase.

\paragraph{Attention v.s. MLP}

For attention blocks, with results shown in Figures~\ref{fig:layerwise-patching}(a) and (c), we observe that the query and instruction tokens tend to influence middle layers, particularly in the case of pairwise prompts. This suggests that the attention mechanism integrates context over these layers to enhance relevance modeling between query and document tokens. Significantly, at layer 15, we observe a pronounced indirect effect (IE = 0.35 for pointwise), potentially attributable to the model beginning to assimilate information relayed from prior positions at the terminal token position—information that may encompass semantic information from both documents and queries, as well as task information from the instructions—thereby transitioning towards the formulation of the final output.

For MLPs, as illustrated in Figures~\ref{fig:layerwise-patching}(b) and (d), the document tokens dominate earlier layers (IE = 0.64 at layer 0 for pairwise), and we conjecture that early MLPs store knowledge or semantic information about documents, which is consistent with the viewpoint of the previous work~\cite{geva2022Transformer, meng2022Locating}. The influence of query and instruction tokens also emerges more prominently in the middle layers, but lower than attention blocks.


In summary, the attention blocks seem pivotal for modeling complex interactions or coping information across tokens, whereas the MLPs contribute more to capturing information from tokens. 

\paragraph{Pointwise v.s. Pairwise}

Furthermore, we can analyze the differences in model behavior across various prompt formats, specifically pointwise and pairwise. A vertical comparison in Figure~\ref{fig:layerwise-patching} reveals a fundamentally similar trend. For instance, when patching the attention output at last token, both pointwise and pairwise prompts exhibit the highest indirect effect at layer 15. Conversely, when patching the MLP output at the document position, both approaches demonstrate the highest indirect effect at layer 0.

\begin{figure}
    \centering
    \includegraphics[width=0.99\linewidth]{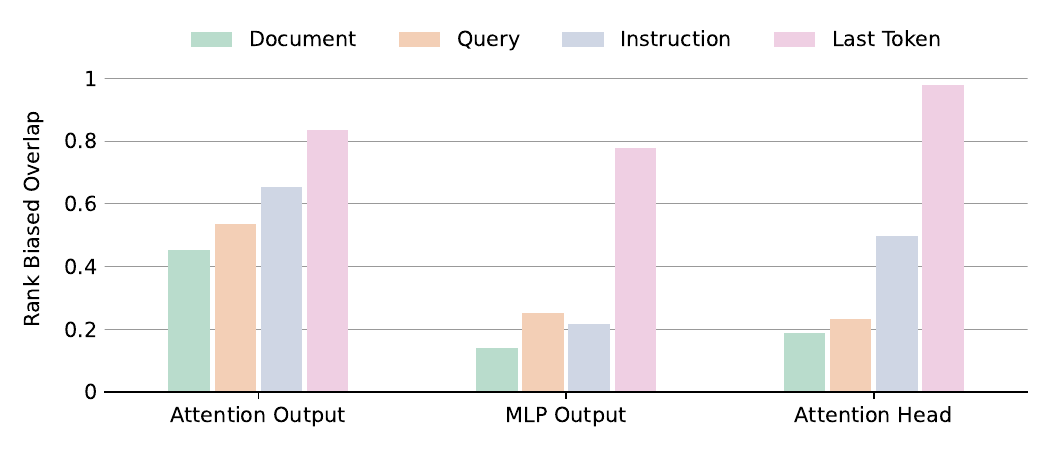}
    \caption{Rank Bias Overlap ($p = 0.7$) of pointwise and pairwise prompts at different components and token positions.}
    \label{fig:rbo}
\end{figure}

To further quantify this similarity, we compute the Rank Bias Overlap (RBO)~\cite{webber2010rbo} of the pointwise and pairwise patching results. Specifically, for a particular module (i.e., Attention or MLP)  and token position  (i.e., document, query, instruction, and last token), we can calculate the indirect effect (IE) at every layer, and rank the layers according to the IE value. Then we can compute the RBO between the two rankings obtained from both pointwise and pairwise methods to measure the similarity. Fully disjoint rankings result in RBO = 0, while identical rankings result in RBO = 1. The results are plots in Figure~\ref{fig:rbo}. From the results, we can see that in terms of attention, the RBO at various positions exceeds 0.5, except for the document position, which suggests a significant similarity between the two rankings. In the case of MLP, only the RBO at the final token position approaches 0.8, while those at other positions are comparatively lower. Nevertheless, as illustrated in Figure~\ref{fig:layerwise-patching}, the influence of the document position at level 0 is paramount for the MLP so that the lower RBO may be attributed to discrepancies in rankings towards the end, rather than reflecting the similarity of rankings at the beginning.

This finding suggests that \textbf{large language models possess a universal mechanism for assessing relevance internally}, applicable both to pointwise evaluations of whether a specific document is relevant to a query and to pairwise assessments of which document is more relevant to the query.

It is important to note that in this paper, we designed a pairwise prompt that is analogous to the pointwise prompt, both of which yield outputs of either ``yes'' or ``no''. In the context of activation patching, we generally assume that the results are based on specified prompt distribution, and thus cannot make statements under different prompts~\cite{heimersheim2024How}. However, in our pilot experiments, we observed that when the model was instructed to output ``document A'' or ``document B'', there were no significant discrepancies in the results from the earlier layers of the model; it was only in the last token position of the later layers that we noted considerable differences. This discrepancy is likely attributable to the different instructions, while the information flow processing of documents and queries remains consistent in the earlier stages. Hence, we can tentatively regard the findings here as reliable.

The above findings reveal how different components at the layer level of the model prioritize various types of input tokens. Given that the output of attention can be expressed as the sum of each attention head, we will subsequently conduct activation patching specifically on the outputs of individual attention heads in the following subsection for further exploration.

\begin{figure}
    \centering
    \includegraphics[width=0.99\linewidth]{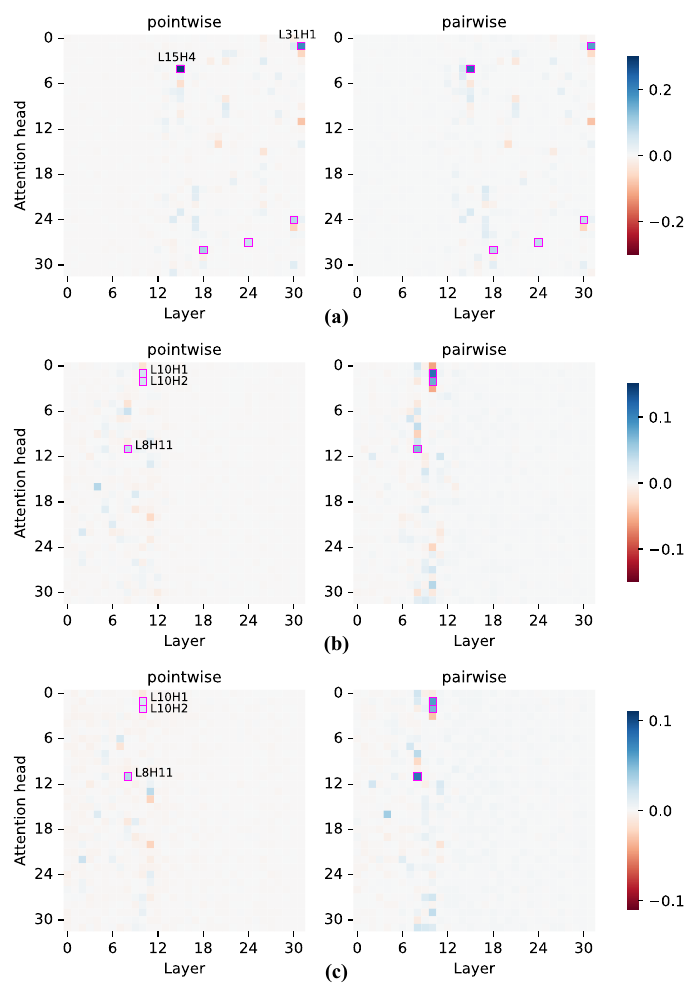}
    \caption{Indirect effect of individual head. (a) Head's output at last token. (b) Heads' output at the position of the query. (c) Heads' attention scores at the position of query-document. Several heads with the highest effects are highlighted in pink.}
    \vspace{-2mm}
    \label{fig:head}
\end{figure}




\subsection{Analysis of Individual Attention Heads}
\label{sec:head}

In this subsection, we focus on the individual attention heads in LLMs and demonstrate how they contribute to relevance judgment.

\paragraph{A few attention heads influence the output format}

We initially investigate which attention heads significantly influence the final output. Figure~\ref{fig:head}(a) illustrates the results of applying activation patching to the outputs of individual attention heads at the last token position. From the figure, it is evident that the attention heads affecting the output are highly sparse, with the effect for the majority of attention heads, particularly in the earlier layers, being nearly zero. Furthermore, we observed that two attention heads, namely L15H4 (Head 4 at Layer 15) and L31H1, exhibit relatively high effects, which are markedly greater than those of other attention heads. This may indicate that these two heads are critical components in controlling the output of ``yes''. 

To verify this, we further check the \emph{unembedding projection} of these two heads by computing $\bm{W}_U \bm{a}^{(h, l)} + \bm{b}$ which is similar to Equation~\eqref{eq:logit} but use the activation output of attention head instead, and observe the top token indices after projection. We observed that ``yes'' (or ``no'') consistently exhibits the highest logit in L31H1, which further corroborates our hypothesis. In contrast, this phenomenon is less pronounced in L15H4, where the tokens with elevated logits appear to be somewhat random. We speculate that L15H4 may serve more as a functional component, guiding the direction of the model's subsequent processing outputs, with the information it conveys requiring further refinement in the subsequent layers before it culminates in the final output.

\paragraph{Several attention heads capture the interaction between the query and document}

Now we have identified the components within the model to fulfill tasks and generate outputs, furthermore, we seek to ascertain whether there are any components employed to process relevance signals. To explore this, we conduct two activation patching experiments, the first patching the output vector of individual attention heads at the position of query, while the second patching the attention scores where the query (i.e., target position) attends to the document (i.e., source position).

\begin{figure}
    \centering
    \includegraphics[width=0.99\linewidth]{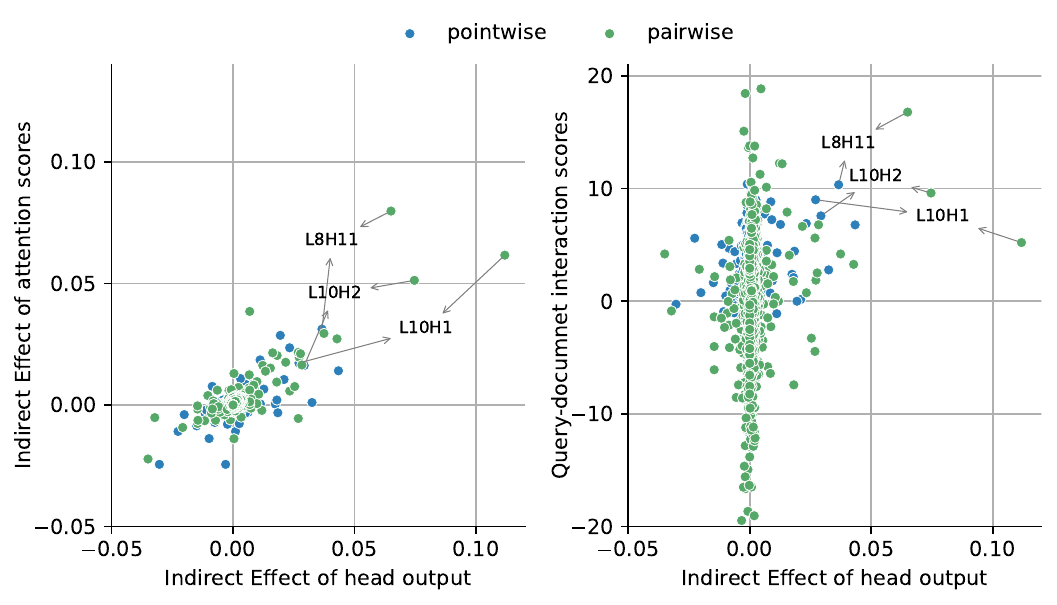}
    \caption{Comparison between head output and attention scores. Left: indirect effect of patching head output at query position vs indirect effect of patching attention scores at query-document position. Right: indirect effect of patching head output at query position vs query-document attention interaction scores.}
    \label{fig:corr}
\end{figure}

The results are shown in Figure~\ref{fig:head}(b) and (c). We can observe that the distribution of the effect the attention head outputs closely resembles that of the attention scores, especially the top heads, e.g., L8H11, L10H1, and L10H2. Additionally, we find a strong correlation between the two effects, as shown in the left part of Figure~\ref{fig:corr}, and the correlation coefficient is 0.65 and 0.82 for pointwise and pairwise respectively. Through this phenomenon, we suggest that within the attention head responsible for modeling and transmitting relevance signals, this modeling (at least part of it) occurs through the interaction between query and document in the attention mechanism, which is subsequently manifested in the output of the attention head.

Furthermore, we hope to use a metric to measure the interaction between query and document. Inspired by the previous studies using the interaction between the query and document for document ranking~\cite{khattab2020ColBERT, chen2024Attention}, we propose computing a heuristic metric, namely \emph{attention interaction score}. First, we define:
\begin{equation}
    s^{(l,j)} (q, d) = \sum_{i \in \mathcal{I}_q} \max_{k \in \mathcal{I}_d} \bm{A}^{(l, j)}_{i,k},
\end{equation}
where $\mathcal{I}_q$ and $\mathcal{I}_d$ represent the set of token indices for query and document respectively, $\bm{A}^{(l, j)}_{i,k}$ denotes the attention weight from the $i$-th token (in the query) to the $k$-th token (in the document) by the $j$-th attention head at layer $l$. Subsequently, we can derive the attention interaction score for each head by calculating the difference between the score obtained using positive samples and that obtained using negative samples:
\begin{equation}
    S^{(l, h)} = \left( s^{(l,h)} (q, d_{pos}) - s^{(l,h)} (q, d_{neg}) \right)
\end{equation}
For pointwise, we perform forward pass on both positive and negative once each. For pairwise, only one forward pass is needed.

We plot the results in the right part of Figure~\ref{fig:corr}, comparing the attention interaction scores with the indirect effect of patching head output. Although the correlation is not so obvious, we can find that those heads with the highest effect (L8H11 and L10H2) still obtain high attention interaction scores. This further confirms our hypothesis that several attention heads are engaged in processing potential correlation signals by modeling the interactions between queries and documents and transmitting them to the outputs of subsequent modules.

\begin{figure}[t]
    \centering
    \includegraphics[width=0.99\linewidth]{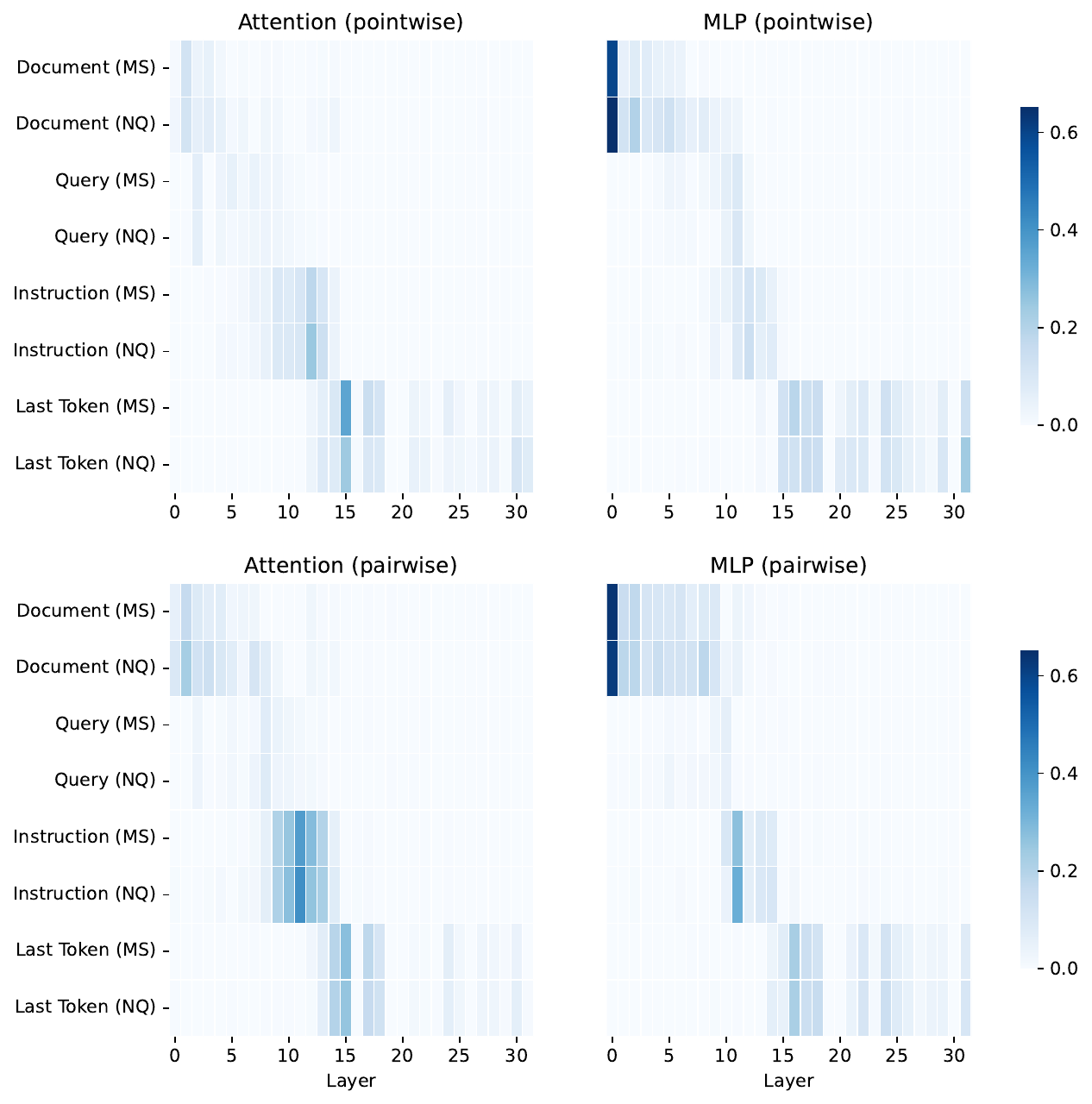}
    \vspace{-3mm}
    \caption{Indirect Effect of different components at different token positions within LLama-3.1-8B-Instruct on both MS MARCO (MS in short) and NQ. The results at the same token positions are placed on adjacent lines for easy comparison.}
    \label{fig:diff_data}
\end{figure}

However, while our experiment identified a correlation between the interaction of queries and documents within the attention mechanism and the output of attention heads at the query position, their influence on the model's final output remains minimal. Even the most prominent attention heads have an indirect effect around 0.1. We postulate two underlying reasons for this phenomenon. First, the input tokens at the query position are entirely identical, which could result in a lack of significant variation in the activations at the corresponding positions, particularly when compared to the elevated indirect effect observed in the early MLP layers at the document position in Figure~\ref{fig:layerwise-patching}(b). Secondly, the relevance signal may not solely arise from the attention interactions between the query and the document; it may also involve semantic knowledge, necessitating the collaboration of multiple modules. Consequently, the influence of a single attention head is somewhat limited.

\subsection{Generalize to Different Datasets and Models}
\label{sec:generalize}

\begin{figure}[t]
    \centering
    \includegraphics[width=0.99\linewidth]{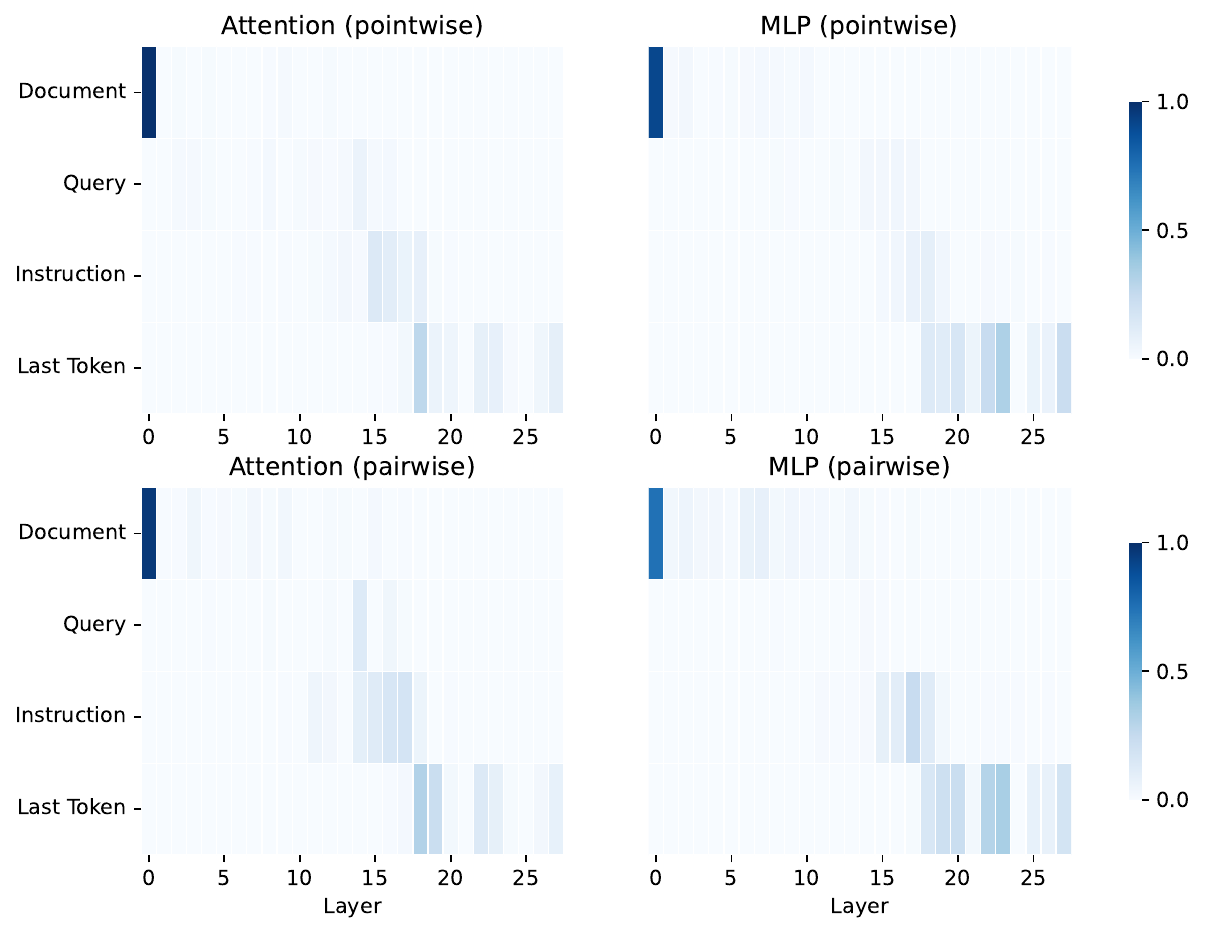}
    \includegraphics[width=0.99\linewidth]{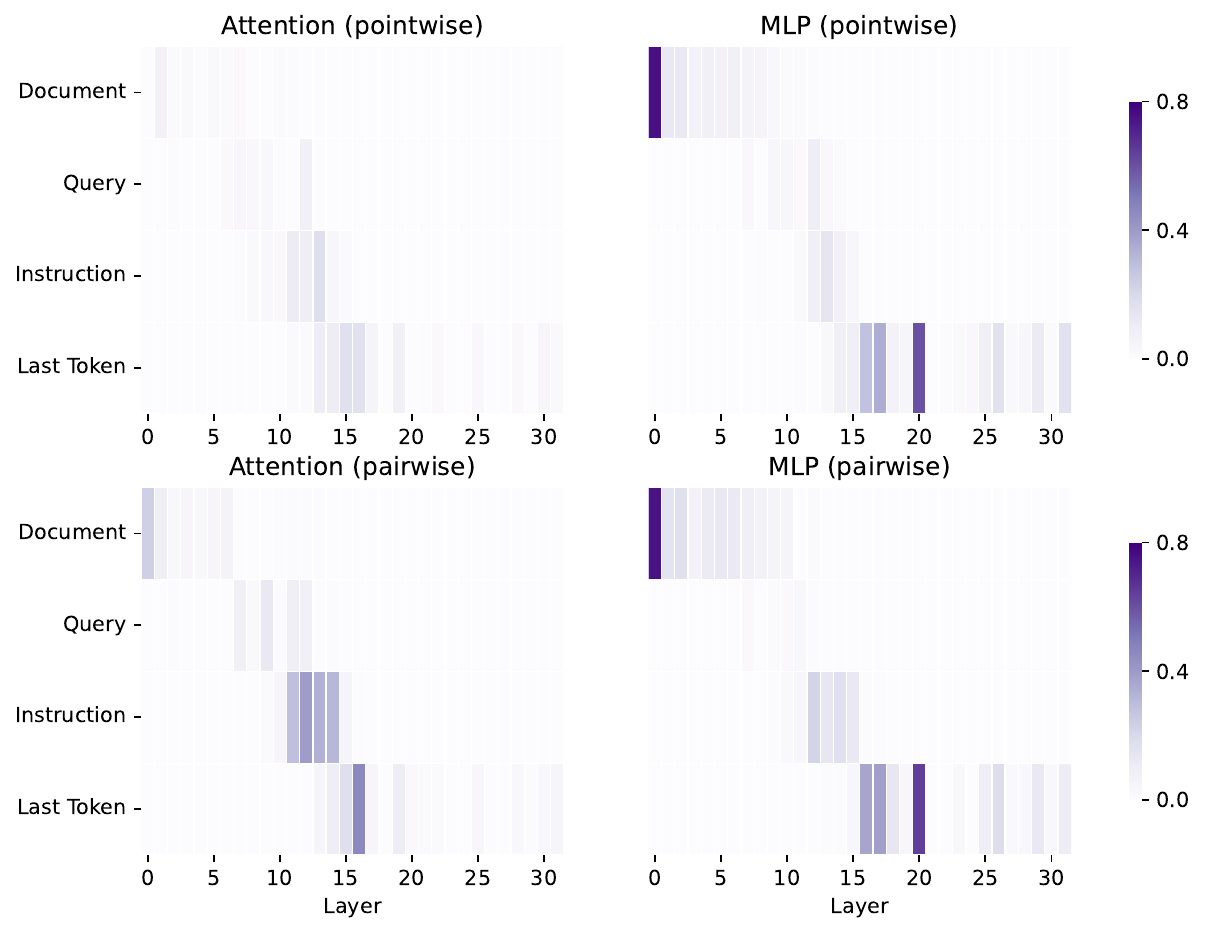}
    \caption{Indirect Effect of different model components at different token positions within Qwen2.5-7B-Instruct (upper in blue) and Mistral-7B-Instruct-v0.3 (lower in purple) on 100 samples from MS MARCO.}
    \label{fig:models}
    \vspace{-2mm}
\end{figure}

\subsubsection{Results on different dataset}

So far, all experiments have been conducted using Llama-3.1-8B-Instruct on a dataset constructed based on MS MARCO. However, there may be different definitions of relevance for different datasets. Therefore, we conduct additional experiments on NQ datasets described in Section~\ref{sec:setup}. Specifically, the experimental procedure is the same as that on MS MARCO, with the results illustrated in Figure~\ref{fig:diff_data}.

From the results, we can see that for different datasets, from low to deep layers, the trend of effect changes is very similar for different components and positions. Based on this, we conclude that there may exists a universal mechanism within the model for modeling relevance, which is independent of data distribution. Regarding the position of documents, the earlier MLPs may have retained semantic information, suggesting the universality of this mechanism, which is quite intuitive. Since instructions are identical, components with a high effect at the instruction position are also identical, which is also reasonable. The crux lies in the interaction mechanism between queries and documents, which may be pivotal for modeling relevance signals. We conjecture that this is facilitated by functional components present within the model, as discussed in Section~\ref{sec:head}, thus remaining independent of the data.

\subsubsection{Results on different model} 

We also conducted experiments using models other than Llama. Specifically, we selected Qwen2.5-7B-Instruct~\cite{qwen2024Qwen25} and Mistral-7B-Instruct-v0.3~\cite{jiang2023Mistral}, two large language models with comparable parameter numbers that similarly excel in relevance judgment tasks. Experiment results are shown in Figure~\ref{fig:models}. Like Llama, we can also observe a progressive manner in these two models and the consistency between pointwise and pairwise. A slight difference is that, both attention and MLP at layer 0 yield high effects at the document position, while effects at the query position are lower. Nevertheless, it's likely that there are similar mechanisms within different large language models for relevance judgment.

\subsection{Evaluation on Downstream Tasks}
\label{sec:eval}

\begin{table}[t]
\small
\caption{Mean ablation results of Llama-3.1-8B-Instruct on relevance judgment (F1-score as the metric) and document reranking (NDCG@10 as the metric, the NDCG@10 of the first-stage retrieval result is 0.51). The performance decrease ratio compared to the full model is indicated in parentheses.}
\label{tab:evaluation}
\begin{tabular}{@{}lllll@{}}
\toprule
     & \multicolumn{2}{c}{\textbf{Relevance Judgment}} & \multicolumn{2}{c}{\textbf{Reranking}} \\
     & Pointwise   & Pairwise          & Pointwise     & Pairwise        \\ \midrule
\textbf{Full model}       & 0.91 & 0.86 & 0.62 & 0.62 \\
 - Random-80          & 0.91 (-0.0\%) & 0.85 (-1.2\%) & 0.60 (-3.2\%) & 0.61 (-1.6\%)\\
 - Doc-20        & 0.90 (-1.1\%) & 0.85 (-1.2\%) & 0.60 (-3.2\%) & 0.60 (-3.2\%)\\ 
 - Query-20           & 0.81 (-11.0\%)& 0.81 (-4.7\%) & 0.58 (-6.5\%) & 0.58 (-6.5\%)\\
 - Inst-20     & 0.78 (-14.3\%)& 0.68 (-20.0\%)& 0.59 (-4.8\%) & 0.57 (-8.1\%)\\
 - Last-20            & 0.55 (-39.6\%)& 0.62 (-27.1\%)& 0.56 (-9.7\%) & 0.55 (-11.3\%)\\
 - Mixed-80           & 0.47 (-48.4\%)& 0.50 (-41.2\%)& 0.51 (-17.7\%)& 0.52 (-16.1\%)\\
\bottomrule
\end{tabular}
\end{table}

In previous sections, we explored which components contribute to relevance judgment and provided a potential workflow explanation within LLMs. Yet we are not sure these observed components are all important in different IR tasks that is related to relevance assessment. To verify this question, we perform evaluation experiments on downstream tasks including relevance judgment and document reranking. Specifically, we use knockout techniques~\cite{wang2022Interpretability} by mean ablating the activation of particular components, i.e., replacing the activation with the mean activation value. Here we conduct evaluation at the level of attention head outputs.

In relevance judgment, we use the previous dataset constructed from MS MARCO and obtain 200 test samples where half are positive. For each query, mean activation is computed independently using one positive example and one negative example. In document reranking, we use TREC DL19 dataset and rerank top-20 documents retrieved by BM25, using pointwise method~\cite{liang2022Holistic} and pairwise method~\cite{qin2023Large}, and mean activation is computed independently using all candidates for each query. The attention heads are selected using two strategies: random sampling and choosing the attention heads with the largest indirect effect.

The results are listed in Table~\ref{tab:evaluation}. Random-80 means we ablate 80 heads at all token positions that are randomly sampled, and the result is taking the average of 5 times. Document-20 means we ablate 20 heads with the highest indirect effect at the document positions, and so forth. Mixed-80 means that we use all four types of top heads listed above. We can see that for both two tasks, ablating several random attention heads will cause almost no performance decrease. However, as shown in the last row of the table, performing ablation at different positions would result in the complete loss of effectiveness of the model in these tasks (an F1-score of 0.5 is equivalent to random classification). This experimental result demonstrates that ablation of less than one-tenth of attention heads can completely render the model ineffective in relevance-related tasks, indicating the necessity of these components we previously discovered in these tasks.

Furthermore, for the results of ablation at four different positions, ablation on the last token has the greatest impact on performance. We believe this is because this position is directly related to the model output. However, the impact of ablation at the document and query positions is not as substantial as we expected, on the contrary, its impact on performance is relatively small. This phenomenon can be explained by the backup behavior~\cite{mcgrath2023hydra} in which the information of query and document is also stored in other components, resulting in a small portion of the components being unable to disrupt the behavior of the model. 

\section{Discussions and Conclusion}

In this paper, we delved into the internal mechanisms by which large language models understand and operationalize relevance. We would like to briefly summarize and highlight several core findings and discuss some potential limitations and future works in this section.

\paragraph{How LLMs perform relevance judgment?}

Through extensive empirical studies using activation patching techniques, we have uncovered several insights into the process of relevance assessment within LLMs, as discussed in Section~\ref{sec:layer}. In general, we found that \textbf{LLMs may process and transmit information in a progressive manner}: early layers capture semantic information from documents and queries, middle layers integrate them with task-specific instructions, and later layers control the output format through specific attention heads. Furthermore, we also discovered that the attention mechanism may play a role in relevance assessment, particularly in modeling the interaction between queries and documents, as shown in Section~\ref{sec:head}.

\paragraph{Is this mechanism universal?}

This question can be viewed from two distinct perspectives. Firstly, the varying prompts, specifically pointwise and pairwise, embody two divergent logics and paradigms for assessing relevance. Our experiment results revealed a notable degree of similarity in model behavior across these two paradigms, as discussed in Section~\ref{sec:layer}. Secondly, regarding different downstream tasks that necessitate the application of relevance, such as relevance judgment and document ranking, our experimental findings indicate that knocking out the same important components adversely affects the performance of both tasks concurrently (see Section~\ref{sec:eval}). These observations suggest that \textbf{the mechanism of assessing relevance within LLMs may serve as a universal mechanism and independent of specific prompts or tasks}. Additionally, experiments on different LLMs show that the observed mechanism is consistent across different models, further confirming the existence of a universal mechanism for relevance assessment as a common feature of LLMs.

\vspace{0.5\baselineskip}

While our study provides valuable insights into the internal workings of LLMs for relevance assessment, several issues warrant further investigation. 
Firstly, in this paper, we employ activation patching techniques to analyze the model; however, it is imperative to acknowledge that, in contrast to previous studies utilizing the same methodology to analyze tasks such as the indirect object identification~\cite{wang2022Interpretability}, the task of relevance judgment presents a greater level of difficulty. This difficulty arises from the diversity of documents and queries, which may encompass various domains of knowledge. Consequently, the analysis presented here is more focused on how the model handles the task of assessing relevance, rather than on the level of semantics and knowledge. We believe that further analysis using different techniques at the knowledge level is worthwhile, but it is beyond the scope of this study.

Secondly, although we identified several crucial components through activation patching, which function at various token positions, and verified their necessity in downstream tasks, we also discovered some of them have somewhat limited effects. For instance, the effect observed at the query position is relatively smaller, and the query position's influence on downstream tasks was less significant compared to the last token. We acknowledge the rationality of this phenomenon; however, the specific roles of these components with diminished effects merit further investigation. Additionally, we cannot assert that the components we have identified are sufficient to accomplish the entire task of relevance judgment; i.e., we cannot claim their sufficiency and minimality. Whether it is possible to identify a minimal circuit that is adequate for completing the task remains a question worthy of exploration.

Finally, as highlighted in Section~\ref{sec:setup}, we refrained from any fine-tuning and utilized off-the-shelf LLMs directly in our experiments. How LLMs acquire the capacity for relevance judgment during either pre-training or post-training remains unexplored. Additionally, as some studies indicate that fine-tuning does not alter but rather enhances the same internal mechanisms~\cite{prakash2024FineTuning}, validating this observation in relevance assessment is also valuable.

In conclusion, this paper represents a primary study toward demystifying the relevance assessment processes within large language models. By employing mechanistic interpretability techniques, we have unraveled the progressive information flow and identified critical pathways that contribute to relevance judgments. Our findings lay the groundwork for enhancing the reliability, fairness, and transparency of LLM-based IR systems. We hope that this research inspires further exploration into the interplay between model architecture and relevance assessment, ultimately fostering the development of more interpretable and trustworthy AI technologies for information retrieval.

\bibliographystyle{ACM-Reference-Format}
\bibliography{reference,addition}

\end{document}